\documentclass[superscriptaddress,twocolumn,showpacs,aps]{revtex4}

\usepackage{graphicx}
\usepackage{dcolumn}
\usepackage{bm}

\renewcommand{\vec}[1]{\mbox{\boldmath $#1$}}

\begin{document}

\preprint{}

\title{
Surface diffuseness anomaly in heavy-ion potentials \\ 
for large-angle quasielastic scattering}

\author{K. Hagino}
\affiliation{
Department of Physics, Tohoku University, Sendai 980-8578, Japan}

\author{T. Takehi}
\affiliation{
Department of Physics, Tohoku University, Sendai 980-8578, Japan}

\author{A.B. Balantekin}
\affiliation{
Department of Physics, Tohoku University, Sendai 980-8578, Japan}
\affiliation{
Department of Physics, University of Wisconsin,
Madison, Wisconsin 53706, USA}

\author{N. Takigawa}
\affiliation{
Department of Physics, Tohoku University, Sendai 980-8578, Japan}

\date{\today}

\begin{abstract}
Recent high precision experimental data for heavy-ion 
fusion reactions at subbarrier energies 
systematically show that a surprisingly 
large surface diffuseness parameter for a Woods-Saxon potential is 
required in order to fit the data. We point out that 
experimental data for quasi-elastic scattering at backward angles 
also favor a similar large value of surface diffuseness parameter. 
Consequently, a double folding approach fails to reproduce 
the experimental excitation function of quasielastic scattering 
for the $^{16}$O + $^{154}$Sm system 
at energies around the Coulomb barrier. 
We also show that the deviation of the ratio of the quasielastic
to the Rutherford cross sections from unity at deep subbarrier
energies offers an unambiguous way to determine 
the value of the surface diffuseness parameter in the nucleus-nucleus 
potential. 
\end{abstract}


\maketitle

The nucleus-nucleus potential is the primary ingredient in 
nuclear reaction calculations. Its nuclear part has  
often been parametrized as a Woods-Saxon form \cite{BW91}. 
Elastic and inelastic scattering are sensitive mainly to the surface 
region of the nuclear potential, where the Woods-Saxon parametrization
has a simple exponential form. This fact has been exploited to 
study the surface property of nuclear potential. Usually, the best 
fit to experimental data for scattering 
is obtained with a diffuseness of 
around 0.63 fm \cite{BW91,CW76,LM80,CPR96,SAC01}. 
This value is consistent with 
a double folding potential\cite{HDG02,GHDN04}, and seems to be well
accepted\cite{BW91,EB96}. 

In marked contrast, recent high precision experimental data for 
heavy-ion fusion reactions at energies around the Coulomb barrier 
suggest that a much larger value of diffuseness, ranging from 0.75
to 1.5 fm, is required to fit the 
data\cite{HDG02,GHDN04,L95,NMD01,HRD03,DHNH04} (See Ref.\cite{NBD04} for 
a detailed systematic study). The Woods-Saxon potential which fits elastic 
scattering overestimates fusion cross sections at energies both 
above and below the Coulomb barrier, having an inconsistent energy 
dependence to the experimental fusion excitation function. When the 
height of the Coulomb barrier is fixed, 
the larger diffuseness parameter leads to 
the smaller barrier position and the smaller barrier curvature (thus the 
larger tunneling region). The main effect on the fusion cross sections 
comes from the barrier position 
and the tunneling width of the barrier at energies above 
and below the Coulomb barrier, 
respectively. A large diffuseness parameter appears to be desirable 
in both these aspects \cite{HDG02}. The reason for the large discrepancies  
in diffuseness parameters extracted from scattering and from fusion 
analyses, however, has not yet been understood.  

The purpose of this paper is to discuss the dependence of quasielastic 
excitation function at a large scattering angle on the surface diffuseness 
parameter in a nucleus-nucleus potential. The quasielastic cross section is 
defined as the sum of the cross sections of 
elastic, inelastic, and transfer reactions. 
Its excitation function at backward angles provides complementary 
information to the fusion process \cite{ARN88,TLD95,HR04}. 
It therefore offers an ideal test ground for a large diffuseness parameter 
suggested by the recent fusion data. 
This is particularly of interest in connection to the steep falloff
phenomena of fusion cross sections at deep subbarrier energies 
observed recently in several systems \cite{JER02,JEB04,JRJ04,HRD03}. This is 
so because the measurement of 
quasielastic scattering is experimentally much easier than that of
fusion reaction, especially at deep subbarrier energies \cite{HR04}. 
Contrary to what one might expect, 
we demonstrate below that the surface diffuseness parameter 
which fits the experimental data of quasielastic scattering 
is consistent with the one for fusion, rather than the commonly 
accepted value for scattering. 

As a concrete example, let us consider the $^{16}$O+$^{154}$Sm reaction. 
Neglecting the finite 
excitation energy of the ground state rotational band
in the target nucleus $^{154}$Sm, the cross sections for 
fusion and quasielastic scattering 
are given by \cite{ARN88,HR04,RHT01,W73} 
\begin{equation}
\sigma_{\rm fus}(E)=\int^1_0d(\cos\theta_T)
\sigma_{\rm fus}(E;\theta_T),
\end{equation}
and
\begin{equation}
\sigma_{\rm qel}(E,\theta)=\int^1_0d(\cos\theta_T)
\sigma_{\rm el}(E,\theta;\theta_T),
\end{equation}
respectively, in the isocentrifugal approximation, where one neglects 
the angular momentum transfer in the centrifugal potential
\cite{HR04,HRK99}. $\theta$ and $\theta_T$ are the scattering
angle and the orientation angle of the deformed target with respect to 
the projectile direction, respectively. 
$\sigma_{\rm fus}(E;\theta_T)$ and $\sigma_{\rm el}(E,\theta;\theta_T)$
are the fusion and the elastic cross sections for the angle dependent 
potential $V(r,\theta_T)$ given by,
\begin{equation}
V(r,\theta_T)=V_N(r,\theta_T)+V_C(r,\theta_T), 
\label{pot}
\end{equation}
\begin{equation}
V_N(r,\theta_T)=\frac{-V_0}{1+\exp[(r-R-R_T\sum_\lambda\beta_\lambda
Y_{\lambda 0}(\theta_T))/a]}, 
\end{equation}
\begin{eqnarray}
V_C(r,\theta_T)&=&\frac{Z_PZ_Te^2}{r}
+\sum_\lambda\left(\beta_\lambda
+\frac{2}{7}\sqrt{\frac{5}{\pi}}\beta_2^2\delta_{\lambda,2}\right)\nonumber \\
&&\,\times\frac{3Z_PZ_Te^2}{2\lambda+1}\frac{R_T^\lambda}{r^{\lambda+1}}
Y_{\lambda0}(\theta_T). 
\end{eqnarray}

\begin{figure}
\includegraphics[scale=0.4,clip]{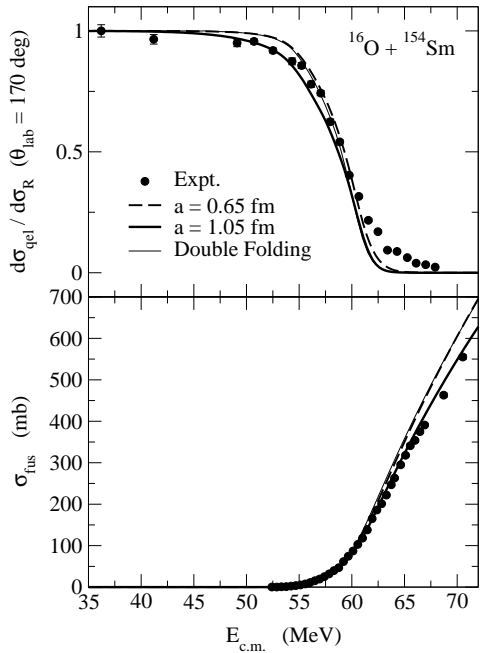}
\caption{
The ratio of quasielastic to the Rutherford cross sections 
at $\theta_{\rm lab}=170$ deg (the upper panel) and the fusion cross 
section (the lower panel) for the $^{16}$O + $^{154}$Sm reaction. 
The solid line is obtained with the orientation-integrated formula
with $\beta_2$=0.306 and $\beta_4$=0.05 by using the Woods-Saxon
potential with the surface diffuseness parameter $a$ of 1.05 fm, while 
the dashed line with $a$ of 0.65 fm. The result of the double
folding potential with the density-dependent 
M3Y interaction is denoted by the
thin solid line. The experimental data are taken from Refs. 
\cite{L95,TLD95}. 
}
\end{figure}

Figure 1 compares the experimental data for the quasi-elastic (given 
as the ratio to the Rutherford cross section; the
upper panel) and the
fusion (the lower panel) 
cross sections with calculated cross sections obtained with different 
values for the surface diffuseness parameter in the Woods-Saxon
potential. The experimental data are taken from Refs.\cite{L95,TLD95}, 
where the quasi-elastic cross sections were measured at 
170 degree in the laboratory frame. 
The solid and dashed lines are 
obtained with a Woods-Saxon potential with 
$a$=1.05 fm and $a$=0.65 fm, respectively. 
The depth and the radius parameters of the potentials are 
$V_0$=165 MeV and $R=0.95\times(A_P^{1/3}+A_T^{1/3})$ fm for the
former, and  
$V_0$=220 MeV and $R=1.1\times(A_P^{1/3}+A_T^{1/3})$ fm for the
latter. 
The deformation parameters 
are taken to be $\beta_2$=0.306 and $\beta_4$=0.05 with $R_T=1.06
\times A_T^{1/3}$ fm. We use a short range imaginary potential 
with $W$=50 MeV, $a_w$=0.4 fm, and $r_w$=1.0 fm in order to simulate 
the compound nucleus formation. The absorption cross sections are thus 
identified with the fusion cross sections. 

It can be clearly seen in the figure that the experimental data favor the 
internuclear potential with the larger value of diffuseness parameter, 
$a$=1.05 fm,  
both for fusion and quasielastic scattering. We have checked that 
the fit to the experimental data with the potential with $a$=0.65 fm 
does not improve 
even if we vary the depth and the radius 
parameters of the potential as well as the deformation parameters. 
The discrepancy between the experimental data and the theoretical 
curve for the quasielastic excitation function around 
$E$=65 MeV is due to the transfer process\cite{TLD95}, which is not 
included in the present calculations. 

\begin{figure}
\includegraphics[scale=0.4,clip]{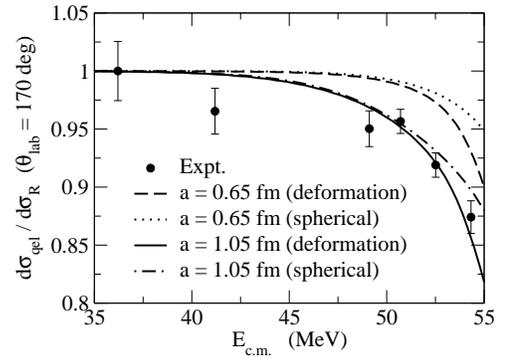}
\caption{
Effects of the deformation of the target nucleus on 
the quasi-elastic scattering for the $^{16}$O + $^{154}$Sm reaction. 
The meaning of the solid and the dashed lines is the same as in 
Fig. 1. The dot-dashed and the dotted lines are obtained by assuming 
a spherical target for 
a Woods-Saxon
potential with a surface diffuseness parameter $a$ of 1.05 fm and 0.65 fm, 
respectively. 
}
\end{figure}

For a single channel problem, 
the ratio of the elastic 
to the Rutherford cross sections at backward angles is given 
by \cite{HR04,LW81}
\begin{equation}
\frac{d\sigma_{\rm el}}{d\sigma_R}(E,\theta)
\sim
1+\frac{V_N(r_c)}{ka}\,
\frac{\sqrt{2a\pi k\eta}}{E},
\end{equation}
at energies well below the Coulomb barrier, where the tunneling 
probability is exponentially small (see Ref.\cite{HR04} for a more 
general formula which is valid also at higher energies). 
This formula is obtained with the semiclassical perturbation theory 
by assuming that the nuclear potential
$V_N(r)$ is proportional to $\exp(-r/a)$ around the distance of closest 
approach, that is, $r_c=(\eta+\sqrt{\eta^2+\lambda_c^2})/k$, 
where $\eta$ is the Sommerferd parameter and
$\lambda_c=\eta\cot(\theta/2)$. 
The deviation of the ratio of the cross sections at subbarrier energies  
from unity is therefore sensitive only to 
the surface property of nuclear potential, and provides 
a relatively model independent way to study the effect of 
surface diffuseness parameter. 
In order to demonstrate that 
the surface diffuseness is indeed more influential than the channel 
coupling effect to quasielastic scattering at low energies, 
Figure 2 shows the effect of 
deformation of the target nucleus on the quasielastic cross sections. 
We find that the effect is negligible at deep subbarrier energies, 
and 
the role played by the 
surface diffuseness parameter is indeed identified unambiguously. 
The strongest energy dependence of the cross 
section ratio comes from the exponential 
factor, $\exp(-r_c/a)$, in the nuclear potential $V_N(r_c$). 
The larger value of 
diffuseness parameter results in the stronger energy dependence, 
and thus the larger deviation of the ratio from unity. 
The measured quasielastic cross sections 
at energies between 35 and 55 MeV are clearly inconsistent 
with $a$=0.65 fm. 
As in subbarrier fusion reactions, a 
larger diffuseness parameter seems to be required in order to fit the
experimental data. 

For completeness of our study, 
we next examine the performance of a double folding potential 
\cite{SL79,BS97,KS00} for the subbarrier reactions. In order to construct 
a nucleus-nucleus potential with the double folding procedure, we assume 
a deformed Fermi function for the (intrinsic) target density, 
\begin{equation}
\rho_T(\vec{r})=
\frac{\rho_0}{1+\exp[(r-R-R\sum_\lambda\beta_\lambda
Y_{\lambda 0}(\hat{\vec{r}}))/a_d]}. 
\label{density}
\end{equation}
We use the same parameters as in Ref. \cite{C76}, including 
the $\beta_2$ and $\beta_4$ deformations. 
We numerically expand Eq. (\ref{density}) into multipoles up to 
$L$=6, and 
construct the double folding potential for each multipole components, leading 
to an orientation dependent potential which corresponds to Eq. (\ref{pot}). 
We use the same (spherical) 
density for $^{16}$O as in Ref.\cite{FS85}. 
For an effective nucleon-nucleon interaction, we use the 
density-dependent Michigan three-range 
Yukawa (DDM3Y) interaction \cite{KBLS84}, together with the zero-range 
approximation for the exchange contribution (See Ref. \cite{BS97} for the 
parameters). We introduce an overall scaling factor to the nuclear potential 
so that the barrier height is the same as that of the Woods-Saxon 
potentials. 
The cross sections computed with the double folding potential thus obtained 
are denoted by the thin solid line in Fig. 1. 
Those are similar to the results of the Woods-Saxon potential with the 
diffuseness parameter of $a$=0.65 fm. In particular, 
compared
with the experimental data, 
the double folding potential leads to a much weaker fall off of quasielastic 
cross sections at energies well below the Coulomb barrier. 
Evidently, the double folding model does not provide a good representation 
both for the quasielastic scattering and the fusion reaction at
subbarrier energies. 

In summary, we have studied the sensitivity of large angle 
quasielastic scattering 
to the surface diffuseness parameter in the nucleus-nucleus potential. 
We have argued that the deviation of 
the ratio of quasielastic to the Rutherford cross sections from unity 
at deep subbarrier energies is sensitive mainly to the surface property 
of nuclear potential, and thus provides 
a useful way to determine the value of surface diffuseness parameter. 
Using this fact, we have shown that the experimental excitation function 
for quasielastic scattering at energies around 
the Coulomb barrier can be reproduced only when a much larger diffuseness 
parameter is used in a Woods-Saxon potential than the commonly accepted 
value, that is, around 0.63 fm. 
This finding is consistent with a recent observation 
in heavy-ion subbarrier fusion reactions. 
It would be helpful to perform other quasi-elastic measurements at 
deep subbarrier energies, so that a systematic 
study for the diffuseness parameter for scattering process is
possible. 

We have also discussed the applicability of a double folding potential 
in quasielastic scattering. We have shown that the cross sections obtained 
with the double folding potential is similar to the one obtained with 
a Woods-Saxon potential whose surface diffuseness parameter is 
around 0.65 fm. Consequently, the double folding potential does not reproduce 
the experimental excitation function for large angle quasielastic 
scattering around the Coulomb barrier. This may appear 
rather surprising, given 
that a double folding approach has enjoyed success in reproducing an angular 
distribution for elastic and inelastic scattering in many systems. 
In order to reconcile this apparent contradiction, 
a more careful investigation, e.g., for the energy 
dependence of a double folding potential due to the exchange contribution 
would be necessary. 
We will report this in a separate paper. 

\bigskip

This work was supported by the Grant-in-Aid for Scientific Research,
Contract No. 16740139, 
from the Japanese Ministry of Education, Culture, Sports, Science and
Technology, 
the U.S. National Science Foundation Grant
No. PHY-0244384, and the University of Wisconsin Research Committee 
with funds granted by the Wisconsin Alumni Research Foundation. 
A.B.B. gratefully acknowledges the 21st Century for Center of Excellence
Program ``Exploring New Science by Bridging Particle-Matter
Hierarchy'' at Tohoku University for
financial support and thanks the Nuclear Theory Group at Tohoku
University for their hospitality.

\end{document}